# COMPARATIVE STUDY OF PROTOCOLS USED FOR ESTABLISHING VPN

P. VENKATESWARI*,  Dr. T. PURUSOTHAMAN**
AP/CSE, Erode Sengunthar Engineering College*, AP/CSE, GCT,Coimbatore **
*  98422 22068,venkat_saran@yahoo.com

**Abstract**
This is an Internet era. Most of the organizations try to establish their development centers and branch offices across the World. Employees working from their homes are also becoming very popular and organizations benefit financially by utilizing less office space, and reducing total expenses incurred by having office workers on site. To meet such requirements   organizations develop a need to communicate with these offices over highly secure, confidential and reliable connections regardless of the location of the office. Here the VPN plays a vital role in establishing a distributed business model.

**Keywords:** VPN, IPSec,SSL,

**Introduction**

In the years gone by if a remote office needed to connect with a central computer or network at company headquarters it meant installing dedicated leased lines between the locations. These dedicated leased lines provided relatively fast and secure communications between the sites, but they were very costly.

To accommodate mobile users companies would have to set up dedicated dial-in remote access servers (RAS). The RAS would have a modem, or many modems, and the company would have to have a phone line running to each modem. The mobile users could connect to the network this way, but the speed was extremely slow and made it difficult to do much productive work.

With the advent of the Internet much of that has changed. If a web of servers and network connections already exists, interconnecting computers around the globe, then why should a company spend money and create administrative headaches by implementing dedicated leased lines and dial-in modem banks?

Well, the first challenge is that we need to be able to choose who gets to see what information. If you simply open up the whole network to the Internet it would be virtually impossible to implement an effective means of preventing unauthorized users from gaining access to the corporate network. Companies spend tons of money to build firewalls and other network security measures aimed specifically at ensuring that nobody from the public Internet can get into the internal network.

How do we reconcile wanting to block the public Internet from accessing the internal network with wanting your remote users to utilize the public Internet as a means of connecting to the internal network? Well we implement a Virtual Private Network (VPN).

 A VPN creates a virtual "tunnel" connecting the two endpoints. The traffic within the VPN tunnel is encrypted so that other users of the public Internet cannot readily view by intercepting communications.

By implementing a VPN, a company can provide access to the internal private network to clients around the world at any location with access to the public Internet. It erases the administrative and financial headaches associated with a traditional leased line wide-area network (WAN) and allows remote and mobile users to be more productive. Best of all, if properly implemented, it does so without impacting the security and integrity of the computer systems and data on the private company network.

**Protocols used to establish a VPN**

A VPN creates a private and secure connection, known as tunnels, through systems that use the data communication capability of an unsecured and public network—the Internet. VPNs use secure protocols to provide private communications over the Internet; they also connect the private corporate network to home office employees, or to a remote business site through virtual connections routed through the Internet. Organizations which decide to use VPNs as their means of secure communication would choose between the more commonly used IPSec and SSL secure protocols. Both protocols have their advantages and disadvantages; the deciding factors between the two depend on the infrastructure of the corporate network, its specific security requirements, costs, and reliability.

Many organizations choose IPSec VPNs through the internet because the cost of private WAN connections, leased lines, and long distance phone charges is extremely high.

Organizations and companies save tremendously by choosing IPSec VPNs. It aids in productivity by increasing business-to-business communications, sales, and customer service management. IPSec provides the capability of allowing home-office employees and telecommuters to connect to the organizations network resources securely and conveniently via remote access through the internet.

**Data link layer**

**Data Link Layer VPNs** can protect various network protocols, so they are often used for non-IP protocols. One type of data link layer VPN is a provisioner-provided VPN, which can protect communications on a dedicated physical link. Data link layer VPNs are most commonly used on top of PPP to secure modem-based connections, although PPP actually encrypts the traffic.

● PPTP protects communications between a PPTP-enabled client and a PPTP-enabled server, and uses GRE to transport data between them.

● L2TP protects communications between an L2TP-enabled client and an L2TP-enabled server, and uses its own tunneling protocol over UDP port 1701 to transport data.

● L2F protects communications between two network devices, such as ISP network access servers and VPN gateways. It is transparent to users, but it does not protect communications between users systems and ISPs.

**Comparison of Protocols Used for Establishing VPN at Data Link Layer**

| Attribute | PPTP | L2TP | L2F |
|---|---|---|---|
| Level of Protection | Protect Non IP Protocol | Protect Non IP Protocol | Protect Non IP Protocol, which is transparent to client |
| Authentication protocol used | - | RADIUS | RADIUS |
| Can be used as an alternate to IPSec | No | Yes can be used to Protect dial-up communications | No |

**Network Layer**

**IPSec**

Internet Protocol Security (IPsec) is a protocol suite for securing Internet Protocol (IP) communications by authenticating and encrypting each IP packet of a data stream.

IPsec also includes protocols for establishing mutual authentication between agents at the beginning of the session and negotiation of cryptographic keys to be used during the session. IPsec can be used to protect data flows between a pair of hosts (e.g. computer users or servers), between a pair of security gateways (e.g. routers or firewalls), or between a security gateway and a host. Figure 1 explains the scenario where IPSec is used .

Before two devices can communicate on an IPSec VPN, they must first agree on the security parameters used during the communication, which is known as establishing a Security Association. The SA specifies the authentication and encryption algorithms to be used, the encryption keys to be used during the session, and how long the keys and the security association are maintained (IPSec VPNs).

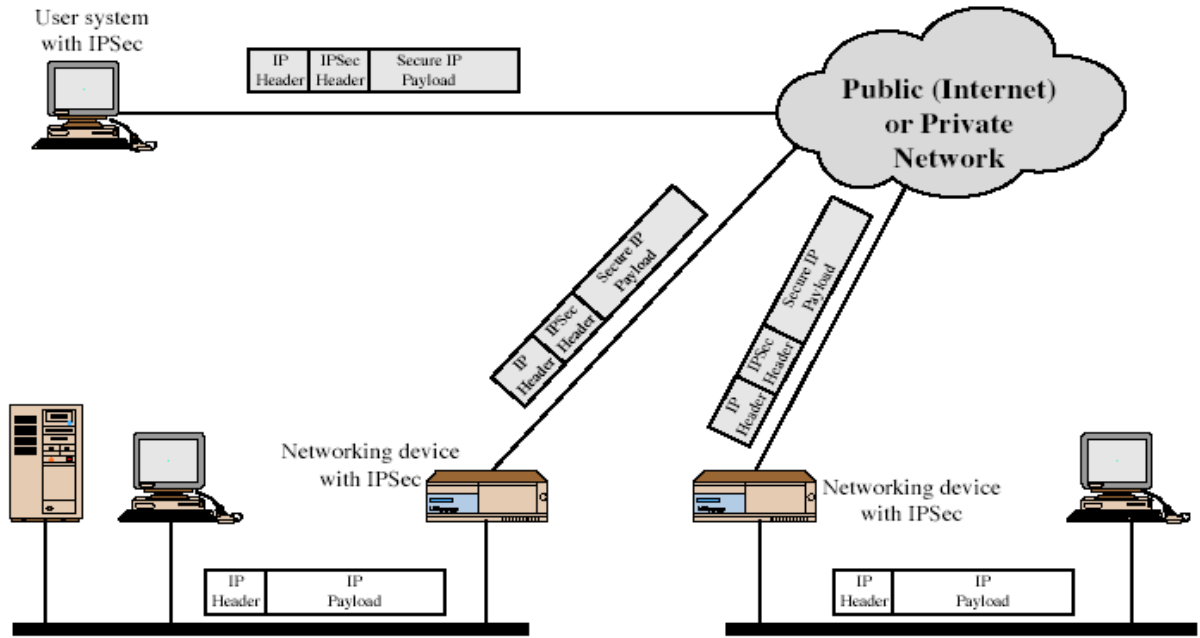

Figure1

IPSec uses two protocols to provide traffic security; one is the Authentication Header (AH) protocol which provides connectionless integrity, data origin authentication, and also an anti-replay service which is optional. The other is the Encapsulating Security Payload (ESP) protocol which provides confidentiality, and limited traffic flow confidentiality. The AH protocol does not provide secrecy for the content of a network communication, but the ESP protocol does along with system authentication and data integrity verification. Either of these protocols can be used in IPv4 or IPv6 and can be used together. If they are used in conjunction with each other, they can provide a wider range of security services. IPSec works in two modes of operation known as transport and tunnel mode. In transport mode only the IP data is encrypted not the IP headers, which allows intermediate nodes to read the source and destination addresses. In tunnel mode, the entire IP packet is encrypted and is then placed as the content portion of another IP packet. Once this is done, these systems then transmit the decrypted packets to their true destinations.

**Application Layer**

**SSL**

Some organizations choose Secure Sockets Layer (SSL) protocol to implement their VPN. SSL is also used to allow access to home-office workers and telecommuters who need to use their organization's resources from outside of the internal network. SSL (VPN) devices allow us to put a device behind the corporate firewall, and basically establish an SSL session from any browser. Basically, these devices will negotiate the session and determine what you'll have access to. Therefore, end users are not required to be at their PC to gain access to the organization's network; they can simply access the organization's network resources via any device which utilizes a web browser which supports SSL sessions.

Secure Sockets Layer (SSL) is a protocol developed by Netscape for securely transmitting documents over the Internet. SSL uses a private key to encrypt data that is transferred over the SSL connection Along with providing data encryption, SSL also provides integrity and server authentication. When SSL is properly configured, it also provides client authentication. SSL works by establishing a normal HTTP session between a client and a server.

**Handshake Protocol Action:**

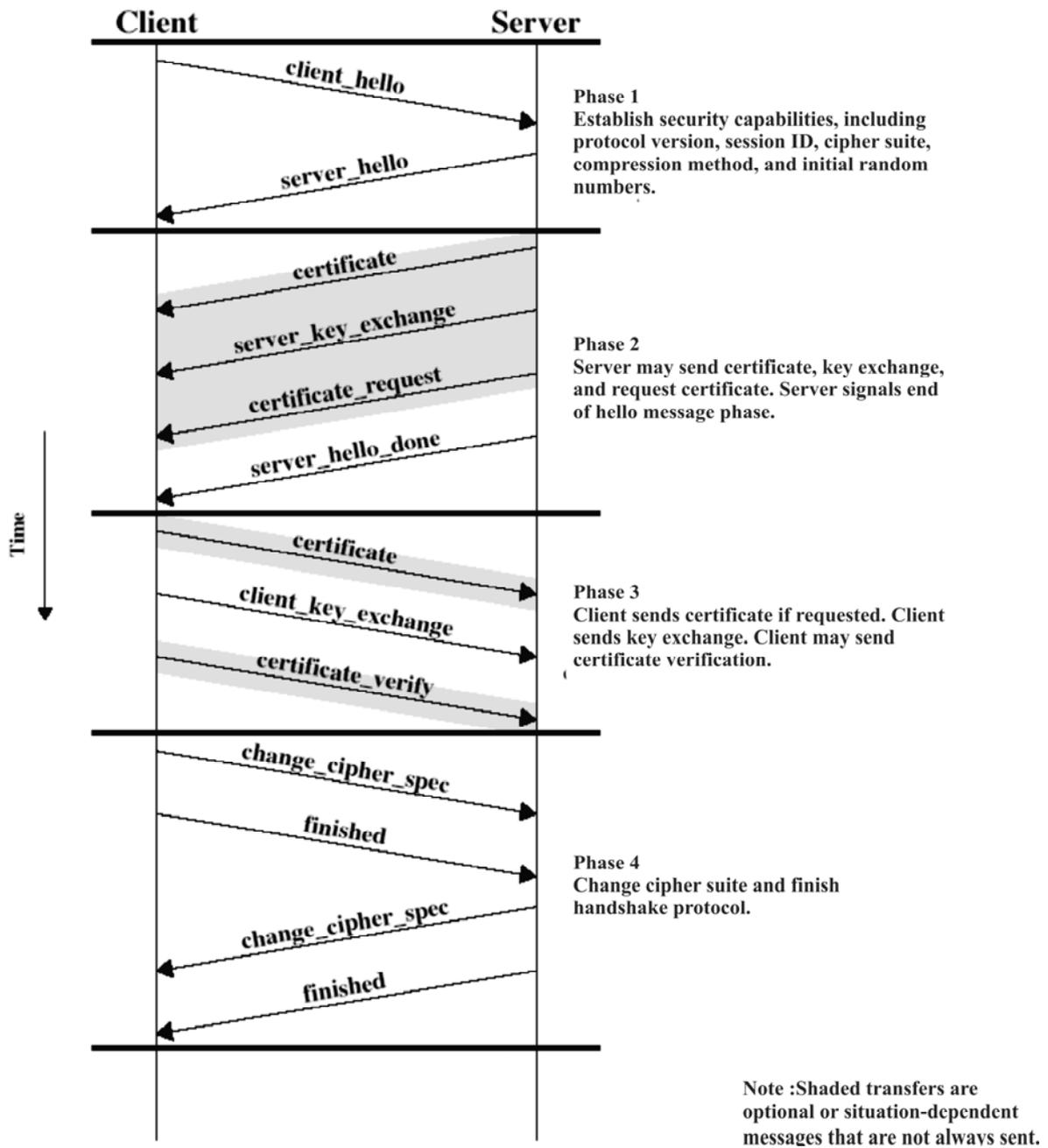

Figure2

Once the client requests for access to a portion of the Web site, which requires secure communications, the server sends a message to the client indicating that a secure connection needs to be established. The client responds with its public key security parameters and the handshaking phase is complete once the server finds a public key match and responds by sending a digital certificate to the client in order to authenticate itself. Once this process is complete, it is up to the client to verify that the certificate is valid and then the SSL session is established between the server and the client. A digital certificate is a public key that has been digitally signed by a recognized authority attesting that the owner of the key is who they say they are. Digital signatures are used to prove that the person sending the message with a public key is actually who he or she claims to be, also proving that the message was not altered, and that it was actually sent .Data can be transmitted securely in any amount, as long as the SSL session remains active between the client and the server. To ensure that data is not altered during transmission, SSL uses cryptographic hashing. Hash values are computed by both the web server and the web browser using the same hash algorithm. If these two values match you can ensure that the data transmitted has not been altered. The pictorial representation of the above explained transactions are given in Figure2.

SSL provides two layers of protocols within the TCP framework; the SSL Record Protocol which is responsible for the fragmentation, compression, encryption, and attachment of an SSL header to the clear text prior to transmission, and Standard HTTP which provides the Internet communication services between client and host, without consideration for encryption of the data that is communicated over the connection .The SSL Record has two parts, the header and the data. The header of a SSL Record can be 2-bytes in length with a maximum record length of 32767 bytes, or the header can be 3-bytes in length with a maximum record length of 16383 bytes.

Comparison between IPSec and SSL

**Table1 given below summaries the comparison between IPSec and SSL Protocols in various attributes**
**Comparison of IPSec and SSL Protocols**

| S. No. | Attribute | IPSec | SSL |
|---|---|---|---|
| 1 | OSI layer | IPSec works on the Network Layer of the OSI Model | SSL works on the Application Layer of the OSI Model |
| 2 | Access Control | All application and resources within a network segment to all users in the VPN. | Controlled access to specific application |
| 3 | Additional Infrastructure | IPSec requires client end software to connect to the VPN. | Content switches with SSL accelerators can encrypt and decrypt data at the network edge, eliminating the need for a Web server's CPU to perform any SSL-related calculations |
| 4 | Optimum choice for | Site to site connection, even remote access | Supports mobile user |
| 5 | Order of applying cryptographic functions | Encrypt first, then calculate MAC on it. At verification it is easy to find any modification that occurs in the middle, by taking MAC on received Cipher text. | MAC is calculated on plaintext then encrypt the plain text . At verification end decryption become mandatory, to identify the modification if any in the middle. |
| 6 | Direction | Bidirectional, Phase –I negotiation completed, both initiator and responder can initiate Phase – II negotiation. | Unidirectional |
| 7 | Multi user support | Multi-user can use one tunnel between two endpoints. Advantage – Lower overhead. | Multi-users have individual connection and different encryption key for each connection. Advantage- Compromising one connection doesn't compromise the other connection |
| 8 | Authentication | Mutual Authentication | Sever Authentication , Client Authentication |
| 9 | Application support | All IP application | Web enabled application |
| 10 | Increase in Packet size(in Bytes) | Maximum – 48Bytes | 25 Bytes |

**Conclusion**

In this paper comparative study has been made for the protocols used for establishing VPN at different layer of OSI model.

Each protocol has its own merits and demerits. It is the choice of the organization to choose the protocol according to their security requirement and business model.

If the organization allows the remote employee to access the office network with official resource, then IPSec VPN is preferable. But it will not support Roaming. If the employee is a member of the marketing team he wants to refer the details in the office network then and there at what ever place he resides at present, even without the company resource, then SSL based VPN is an optimum choice.

In some occasions if the employee wants to have the long session with the desktop available in the office, then IPSec based VPN is the best choice.

Or else, if the employee is on the move he need only transient connectivity with the head office, then second option be better.